\documentclass[aps,prd,onecolumn,eqsecnum,amsmath,nofootinbib,preprintnumbers]{revtex4}%

\usepackage{color,graphicx,float,subfigure}
\usepackage{amsfonts,amssymb,theorem,mathrsfs,times}
\usepackage{bm}
\usepackage{mathtools}
\usepackage{amsfonts,amssymb,theorem,mathrsfs}
\usepackage{dsfont}
\usepackage{setspace}

{\theorembodyfont{\upshape}
}
{\theorembodyfont{\upshape}
}
{\theorembodyfont{\upshape}
}
{\theorembodyfont{\upshape}
}
{\theorembodyfont{\upshape}
}
{\theorembodyfont{\upshape}
}

\newcommand{\dalm}{\kern1pt\vbox{\hrule height 0.9pt\hbox{\vrule width
0.9pt\hskip 2.5pt\vbox{\vskip 5.5pt}\hskip 3pt\vrule width
0.3pt}\hrule height 0.3pt}\kern1pt}

\begin{document}
\preprint{\hfill {\small {ICTS-USTC/PCFT-21-11}}}
\title{On  the  ``Einstein-Gauss-Bonnet Gravity in Four Dimension" }

%

\author{ Li-Ming Cao$^{a\, ,b}$\footnote{e-mail
address: caolm@ustc.edu.cn}}

\author{ Liang-Bi Wu$^b$\footnote{e-mail
address: liangbi@mail.ustc.edu.cn}}

\affiliation{$^a$Peng Huanwu Center for Fundamental Theory, Hefei, Anhui 230026, China}

\affiliation{${}^b$
Interdisciplinary Center for Theoretical Study and Department of Modern Physics,\\
University of Science and Technology of China, Hefei, Anhui 230026,
China}


\date{\today}

\begin{abstract}
	To ensure the existence of a well defined linearized gravitational wave equation, we show that the spacetimes in the so-called ``Einstein-Gauss-Bonnet gravity in four dimension" have to be locally conformally flat.
\end{abstract}
\maketitle

 

\section{introduction}

Recently, a novel Einstein-Gauss-Bonnet gravity in four dimension has been proposed~\cite{Glavan:2019inb}. The action of the theory is based on 
\begin{equation}
S=\int d^Dx\sqrt{-g}\left(R-2\Lambda+\frac{\hat{\alpha}}{D-4}L_{\text{GB}}\right)\, ,
\end{equation}
where $R$ is the $D$-dimensional Ricci scalar,  and $\Lambda$ is the cosmological constant. The Gauss-Bonnet term is given by
\begin{eqnarray}
	L_{\text{GB}}=R^2-4R_{\mu\nu}R^{\mu\nu}+R_{\mu\nu\rho\sigma}R^{\mu\nu\rho\sigma}\, .
\end{eqnarray}
Here, the usual Gauss-Bonnet coupling constant $\alpha$ has been replaced by $\hat{\alpha}/(D-4)$.
Then the so-called four dimensional Einstein-Gauss-Bonnet theory is defined by considering the  limit $D\to4$. In Ref.\cite{Glavan:2019inb}, it is also claimed that under this limit the Gauss-Bonnet invariant gives rise to non-trivial contributions to gravitational dynamics. This is quite different from the content of the well-known Lovelock theorem~\cite{Lovelock:1972vz}.  In view of the importance, it is necessary to investigate the theory from various perspectives. For example, the  well-posedness of the initial value problem of the theory.

To investigate the local well-posedness of the initial value problem of a gravitational theory, one should pay attention to the principle symbol of the linearized perturbation equation~\cite{Reall:2021voz, Reall:2014pwa, Papallo:2017qvl, Reall:2014sla}. Consider the equations of motion linearized around a background solution, we  get
\begin{eqnarray}
	P^{\mu\nu\rho\sigma\alpha\beta}\partial_\alpha\partial_\beta\delta g_{\rho\sigma}+\cdots=0\, ,
\end{eqnarray}
where the ellipses denotes the terms with lower than 2-derivatives acting on $\delta g_{\mu\nu}$. Let $\xi_\mu$ be an arbitrary covector, then $P^{\mu\nu\rho\sigma\alpha\beta}\xi_{\alpha}\xi_{\beta}$ is the so-called the principle symbol. 
Hyperbolicity and causality of a gravitational theory is determined by the principal symbol, so it is natural to require that the principle symbol should have a well defined behavior in this novel Einstein-Gauss-Bonnet gravity.  Otherwise, we have no
well defined linearized gravitational wave equations.

\section{Principle symbol}
Suppose we have a $D$-dimensional spacetime. Based on the above ideas, the principal symbol should be well defined under the limit $D\rightarrow 4$. 
For a diffeomorphism invariant theory,  based on the discussion of symmetry of the tensor $P^{\mu\nu\rho\sigma\alpha\beta}$, Reall has proved that  the principle symbol has a form~\cite{Reall:2021voz},
\begin{eqnarray}
\label{P}
	P^{\mu\nu\rho\sigma}(\xi)=C^{\mu(\rho|\alpha\nu|\sigma)\beta}\xi_{\alpha}\xi_{\beta}\, ,
\end{eqnarray}
where for Einstein-Gauss-Bonnet gravity in $D>4$ dimension we have
\begin{eqnarray}
\label{Ctensor}
	C_{\mu\rho\alpha}{}^{\nu\sigma\beta}=\frac{1}{2}\delta_{\mu\rho\alpha}^{\nu\sigma\beta}+\hat{\alpha}\Big[\frac{D+1}{(D-1)(D-2)}\delta_{\mu\rho\alpha}^{\nu\sigma\beta}R-\frac{2}{D-2}A_{\mu\rho\alpha}{}^{\nu\sigma\beta}+\frac{1}{2(D-4)}W_{\mu\rho\alpha}{}^{\nu\sigma\beta}\Big]\, .
\end{eqnarray}
In the above equation, $\delta_{\mu\rho\alpha}^{\nu\sigma\beta}$ is a generalized Kronecker-delta tensor, while $A_{\mu\rho\alpha}{}^{\nu\sigma\beta}$ and $W_{\mu\rho\alpha}{}^{\nu\sigma\beta}$ are defined as follows
\begin{equation}
	A_{\mu\rho\alpha}{}^{\nu\sigma\beta}=
	\left|\begin{array}{ccc}
		R_{\mu}^{\nu} & R_{\mu}^{\sigma} & R_{\mu}^{\beta}\\
		\delta_{\rho}^{\mu} & \delta_{\rho}^{\sigma} & \delta_{\rho}^{\beta}\\
		\delta_{\alpha}^{\nu} & \delta_{\alpha}^{\sigma} & \delta_{\alpha}^{\beta}
	\end{array}\right|+
	\left|\begin{array}{ccc}
		\delta_{\mu}^{\nu} & \delta_{\mu}^{\sigma} & \delta_{\mu}^{\beta}\\
		R_{\rho}^{\mu} & R_{\rho}^{\sigma} & R_{\rho}^{\beta}\\
		\delta_{\alpha}^{\nu} & \delta_{\alpha}^{\sigma} & \delta_{\alpha}^{\beta}
	\end{array}\right|+
	\left|\begin{array}{ccc}
		\delta_{\mu}^{\nu} & \delta_{\mu}^{\sigma} & \delta_{\mu}^{\beta}\\
		\delta_{\rho}^{\mu} & \delta_{\rho}^{\sigma} & \delta_{\rho}^{\beta}\\
		R_{\alpha}^{\nu} & R_{\alpha}^{\sigma} & R_{\alpha}^{\beta}
	\end{array}\right|\, ,
\end{equation}
and 
\begin{eqnarray}
	W_{\mu\rho\alpha}{}^{\nu\sigma\beta}=36\delta_{[\mu}{}^{[\nu}C_{\rho\alpha]}{}^{\sigma\beta]}\, ,
\end{eqnarray}
where $R_{\mu\nu}$ is the Ricci tensor of the spacetime,  and $C_{\rho\alpha\sigma\beta}$ is the Weyl tensor. 
It is not hard to find that $W_{\mu(\rho|\alpha\nu|\sigma)\beta}$ is vanished when $D=4$. 
So  the  last term in the righthand of  Eq.(\ref{Ctensor}) is $0/0$ type when $D=4$. To ensure the existence of the limit $D\rightarrow 4$, we have to impose a condtion
\begin{eqnarray}
\label{W=0}
	W_{\mu(\rho|\alpha\nu|\sigma)\beta}=0\, ,
\end{eqnarray} 
for all $D>4$.  However, in the case with $D>4$, 
by contracting the two indices $\rho$ and $\sigma$ in Eq.(\ref{W=0}), we obtain 
\begin{eqnarray}
\label{C=0}
	C_{\mu\alpha\nu\beta}=0\, .
\end{eqnarray}
This means that the spacetimes of the theory have to be locally conformally flat. Recently,  based on the equations of motion, people have realized that the theory is well defined only when 
\begin{eqnarray}
\label{L}
	\mathcal{L}_{\mu\nu}=C_{\mu\alpha\beta\gamma}C_{\nu}{}^{\alpha\beta\gamma}-\frac{1}{4}g_{\mu\nu}C_{\alpha\beta\gamma\delta}C^{\alpha\beta\gamma\delta}
\end{eqnarray}
is  vanished for  $D>4$, and a well defined four dimensional Einstein-Gauss-Bonnet theory can be generated as $D\to4$~\cite{Gurses:2020rxb, Gurses:2020ofy, Arrechea:2020gjw, Arrechea:2020evj}. 
Here, with a similar logic, we obtained a more restrictive condition, i.e., Eq.(\ref{C=0}), on the metrics of the theory.

\section{Examples}

To make the problem more transparent, let us consider a spacetime manifold $\mathcal{M}^D\cong M^m\times N^n$ with a metric
\begin{equation}
\label{metric}
g_{\mu\nu}dx^\mu dx^\nu=g_{ab}(y)dy^ady^b+r^2(y)\gamma_{ij}(z)dz^idz^j\, ,
\end{equation}
where $D=m+n$, and the coordinates $\{x^\mu\}$ is given by $\{y^1\, ,\cdots y^m\, ;z^1\, ,\cdots\, ,z^n\}$.  The tuple $(M^m\, ,g_{ab})$ forms a $m$-dimensional Lorentzian manifold if $m>1$, and $(N^n,\gamma_{ij})$ is an $n-$dimensional Riemann manifold. 
This Riemann manifold $(N^n,\gamma_{ij})$ is assumed to be an Einstein manifold, i.e.,
\begin{eqnarray}
	\hat{R}_{ij}=(n-1)K\gamma_{ij}\, ,
\end{eqnarray}
where  $\hat{R}_{ij}$ is the Ricci tensor of $(N^n,\gamma_{ij})$, and $K$ is the sectional curvature of the space. The metric compatible covariant derivatives associated with $g_{\mu\nu}$, $g_{ab}$, and $\gamma_{ij}$ are denoted by $\nabla_{\mu}$, $D_a$, and $\hat{D}_i$, respectively. 
For the metric (\ref{metric}), the tensor perturbation can be put into a form~\cite{Cao:2021sty}
\begin{eqnarray}
\label{master1}
\Big(P^{ab}{}_{ij}{}^{kl} D_aD_b + P^{mn}{}_{ij}{}^{kl} \hat{D}_m\hat{D}_n + P^{a}{}_{ij}{}^{kl} D_a + V_{ij}{}^{kl}\Big)\Big(\frac{h_{kl}}{r^2}\Big) = -\frac{2\kappa_D^2}{r^2}\delta T_{ij}\, ,
\end{eqnarray}
where
\begin{eqnarray}
\label{eq1}
P^{ab}{}_{ij}{}^{kl} = P^{ab} \delta_i{}^k\delta_j{}^l -\frac{4\alpha}{r^2}g^{ab} \hat{C}_i{}^k{}_j{}^l \, ,
\end{eqnarray}
\begin{eqnarray}
\label{eq2}
P^{a}{}_{ij}{}^{kl} = P^{a} \delta_i{}^k\delta_j{}^l -4\alpha(n-2)\frac{D^ar}{r} \frac{\hat{C}_i{}^k{}_j{}^l}{r^2}  \, ,
\end{eqnarray}
\begin{eqnarray}
\label{eq3}
P^{mn}{}_{ij}{}^{kl} = P^{mn} \delta_i{}^k\delta_j{}^l + \frac{4\alpha}{r^2}\big(\hat{C}_j{}^{knl} \delta_i{}^m+\hat{C}_i{}^{knl} \delta_j{}^m+\hat{C}_j{}^{mln} \delta_i{}^k +\hat{C}_i{}^{mln} \delta_j{}^k- \hat{C}^{mknl}\gamma_{ij} - \hat{C}_i{}^k{}_j{}^l\gamma^{mn} \big) \, ,
\end{eqnarray}
and
\begin{eqnarray}
\label{eq4}
V_{ij}{}^{kl} &=& V \delta_i{}^k\delta_j{}^l + \frac{2 \hat{C}_i{}^k{}_j{}^l}{r^2} +\alpha \Bigg\{ 4\Bigg[{}^m\!{R} - 2(n-3)\frac{{}^m\!\Box r}{r}  + (n^2 - 7n + 16)\frac{K}{r^2} - (n-3)(n-4)\frac{(Dr)^2}{r^2}\Bigg] \frac{ \hat{C}_i{}^k{}_j{}^l}{r^2}\nonumber\\
&&-\frac{8}{r^4} \hat{C}_{imjn}\hat{C}^{mknl} + \frac{4}{r^4} \hat{C}_{mn j}{}^k{}\hat{C}^{mn}{}_{i}{}^l -\frac{2}{r^4}\hat{C}^{mnpl}\hat{C}_{mnp}{}^k\gamma_{ij }+\frac{\hat{C}^{mnpq}\hat{C}_{mnpq}}{r^4}\delta_i{}^k\delta_j{}^l \Bigg\} \, .
\end{eqnarray}
In the above equations, $\alpha = \hat{\alpha}/(D-4)$, and we have defined
\begin{eqnarray}
\label{eq5}
P^{ab}= g^{ab} + 2(n-2) \alpha\left\lbrace 2\frac{D^aD^br}{r}+\left[(n-3)\frac{K-(Dr)^2}{r^2}-2\frac{\prescript{m}{}{\Box}r}{r}\right]g^{ab}\right\rbrace-4\alpha\cdot {}^m\! G^{ab}\, ,
\end{eqnarray}
\begin{eqnarray}
\label{eq6}
P^{mn}=\Bigg\{1
+2 \alpha\left[ {}^m\!{R}- \frac{2(n-3) {}^m\!{\Box}r}{r}+ (n-3)(n-4)\frac{K-(Dr)^2}{r^2}\right] \Bigg\}\frac{\gamma^{mn }}{r^2}\, ,
\end{eqnarray}
\begin{eqnarray}
\label{eq7}
P^{a}&=&n\frac{D^ar}{r} +  2(n-2) \alpha \Bigg\{4\frac{D^aD^br}{r}+\Big[ {}^m\!{R}
-2(n-1)\frac{ {}^m\!{\Box}r}{r}\nonumber\\ 
&&+(n-2) (n-3)\frac{K-(Dr)^2}{r^2}\Big]g^{ab}\Bigg\}\frac{D_br}{r}- 8\alpha\cdot {}^m\! G^{ab}\frac{D_br}{r}\, ,
\end{eqnarray}
and
\begin{eqnarray}
\label{eq8}
V&=&{}^m\!{R}-2(n-1)\frac{{}^m\! {\Box}r}{r}+\frac{n(n-3)K}{r^2}-\frac{(n-1)(n-2)(Dr)^2}{r^2}-2 \Lambda\nonumber\\
&+&\alpha\Bigg\{{}^m\! L_{GB}  + 8(n-1)\cdot {}^m\! G^{ab}\frac{D_aD_br}{r} -4(n-1)(n-2)\frac{(D^aD^br)(D_aD_br)}{r^2}\nonumber\\
&+& 4(n-1)(n-2)\left(\frac{{}^m\! {\Box}r}{r}\right)^2+2n(n-3)\frac{K\cdot {}^m\! R}{r^2}-2(n-1)(n-2)\frac{(Dr)^2\cdot {}^m\! {R}}{r^2}\nonumber\\
&-&4n(n-3)^2\frac{K\cdot {}^m\! {\Box}r}{r^3}+4(n-1)(n-2)(n-3)\frac{(Dr)^2\cdot{}^m\! {\Box}r}{r^3}\nonumber\\
&-&2n(n-3)^2(n-4)\frac{K\cdot(Dr)^2}{r^4}+(n-3)(n-4)(n^2-3n-2)\frac{K^2}{r^4}\nonumber\\
&+&(n-1)(n-2)(n-3)(n-4)\left[\frac{(Dr)^2}{r^2}\right]^2\Bigg{\}}\, .
\end{eqnarray}
In the above equations, ${}^m\! {\Box}=g^{ab}D_aD_b$ is the Laplace-Beltrami operator on $(M^m\, ,g_{ab})$,   ${}^m\!{R}$ and ${}^m\! G^{ab}$ are the Ricci scalar and Einstein tensor of $(M^m\, ,g_{ab})$ respectively,  and ${}^m\! L_{GB} $ is the Gauss-Bonnet term which is nontrivial only when $m>3$.   The symbol $\hat{C}_{ijkl}$ is the Weyl tensor of the Riemann manifold $(N^n,\gamma_{ij})$.  

First, let us consider the case $m=1$, i.e.,   $D=1+n$, and the metric~(\ref{metric}) becomes 
$$g_{\mu\nu}dx^{\mu}dx^{\nu} = - dt^2 + r^2(t)\gamma_{ij}dz^idz^j\, .$$
In this case, the non-trivial components of the Weyl tensor of this spacetime can be written as
\begin{eqnarray}
\label{C}
C_{ijkl}=r^2\hat{C}_{ijkl}\, ,
\end{eqnarray}
Obviously, from Eqs.(\ref{eq1}), (\ref{eq2}), (\ref{eq3}), and (\ref{eq4}), the limit $D\rightarrow 4$ or $n\rightarrow 3$, can be done only in the case
$\hat{C}_{ijkl}=0$. This means the Weyl tensor the spacetime is also vanished according to the relation (\ref{C}). 
Under this limit, from  Eqs.(\ref{eq5}), (\ref{eq6}), (\ref{eq7}), and (\ref{eq8}), $P^{tt}$, $P^{mn}$, $P^t$, $V$  have following forms
\begin{eqnarray}
\label{Ptt}
P^{tt}= -\Big[1+2\hat{\alpha}\Big(\frac{K}{r^2}+H^2\Big)\Big]\, ,
\end{eqnarray}
\begin{eqnarray}
	\label{Pmn}
	P^{mn}=\Big[1+2\hat{\alpha}\Big(2\dot{H}+H^2-\frac{K}{r^2}\Big)\Big]\frac{\gamma^{mn}}{r^2}\, ,
\end{eqnarray}
\begin{eqnarray}
\label{Pt}
P^{t}=-3H+2\hat{\alpha}\Big[-2\dot{H}-3H^2-\frac{K}{r^2}\Big]H\, ,
\end{eqnarray}
and
\begin{eqnarray}
	\label{V}
	V=4\dot{H}+6H^2-2\Lambda+2\hat{\alpha}\Big[4H^2\dot{H}+3H^4+\frac{K^2}{r^4}\Big]\, .
\end{eqnarray}
where ``$\cdot$" denotes the derivative with respect to the coordinate $t$, and $H=\dot{r}/r$ is Hubble parameter. In the case $K=0$, with the above results,   Eq.(\ref{master1}) exactly reduces to the relevant part in~\cite{Glavan:2019inb}. When matter field
is absent, Eq.(\ref{master1})  gives the gravitational wave equations on vacumm. For example, the gravitational wave equations on Minkowski spacetime, de Sitter spacetime, and anti de Sitter  spacetime.

%

Second, we consider  the case of $m=2$. 
%
The components of the Weyl tensor of the spacetime can be written as~\cite{Cai:2013cja}
\begin{eqnarray}
\label{Weyltensor2}
C_{abcd} &= & 2 c_1 w g_{a[c} g_{d]b}\, ,\nonumber\\
C_{iajb} &=& - c_2 w r^2 g_{ab}\gamma_{ij}\, ,\nonumber\\
C_{ijmn} &=& 2 c_3 w r^4 \gamma_{i[m} \gamma_{n]j} + r^2 \hat{C}_{ijmn}\, ,
\end{eqnarray}
where 
\begin{equation}
w = {}^2\!{R}+2 \frac{\prescript{2}{}{\Box}r}{r}+2 \frac{K-(Dr)^2}{r^2}\, ,
\end{equation}
and
\begin{equation}
c_1= \frac{n-1}{2(n+1)}\, ,\quad c_2 = \frac{n-1}{2n(n+1)}\, ,\quad c_3 =\frac{1}{n(n+1)}\, .
\end{equation}
For the metric (\ref{metric}) with $\hat{C}_{ijkl}=0$ (this is also one necessary condition of the existence of the limit $D\rightarrow 4$),  the tensor perturbation can be put into a form
\begin{eqnarray}
\label{PE1}
\Big(P^{ab} D_aD_b + P^{mn} \hat{D}_m\hat{D}_n  + P^a D_a  + V \Big)\Big(\frac{h_{ij}}{r^2}\Big) = -\frac{2\kappa_{D}^2}{r^2} \delta T_{ij}\, .
\end{eqnarray}
Here,  $P^{ab}$, $P^{mn}$, $P^a$, $V$  have the same  forms as in Eqs.(\ref{eq5}), (\ref{eq6}), (\ref{eq7}), and (\ref{eq8}) except that ${}^m\!G_{ab}=0$ and ${}^m\!L_{GB}=0$ when $m=2$.
It is easy to find that  $P^{ab}$ and $P^a$ have well defined limits under $D\rightarrow 4$ or $n\rightarrow 2$.  However,  generally, $P^{mn}$ is degenerate under this limit.  Actually, we have
\begin{eqnarray}
\label{PijQ2}
P^{mn}=\Bigg\{1
+2 \alpha\left[w- \frac{2(n-2)\prescript{2}{}{\Box}r}{r}+ (n-2)(n-5)\frac{K-(Dr)^2}{r^2}\right] \Bigg\}\frac{\gamma^{mn}}{r^2}\, ,
\end{eqnarray}
To ensure the regularity of ``effective metric" $P^{\mu\nu}=(P^{ab}\, ,P^{mn})$ of the tensor perturbation equation, we have to impose a condition
$w=0$. From Eqs.(\ref{Weyltensor2}), this implies that the spacetime has to be (locally) conformally flat. 

Furthermore, the function $V$ (one part of the effective potential of the theory) can be rewritten as
%
\begin{eqnarray}
\label{V123}
V&=&{}^2\! {R}-2(n-1)\frac{{}^2\!{\Box}r}{r}+\frac{n(n-3)K}{r^2}-\frac{(n-1)(n-2)(Dr)^2}{r^2}-2 \Lambda\nonumber\\
&+&\alpha(n-2 )\Bigg\{-4(n-1)\frac{(D^aD^br)(D_aD_br)}{r^2}+4(n-1)\left(\frac{{}^2\!{\Box}r}{r}\right)^2\nonumber\\
&+&2(n-1)\frac{K\cdot {}^2\!R}{r^2}-2(n-1)\frac{(Dr)^2\cdot {}^2\!{R}}{r^2}-4(n^2-4n +1)\frac{K\cdot {}^2\! {\Box}r}{r^3}\nonumber\\
&+&4(n-1)(n-3)\frac{(Dr)^2\cdot{}^2\!{\Box}r}{r^3}-2(n^3 - 8 n^2 + 17 n - 2)\frac{K\cdot(Dr)^2}{r^4}\nonumber\\
&+&(n^3 - 8 n^2 + 15 n +8)\frac{K^2}{r^4}+(n-1)(n-3)(n-4)\left[\frac{(Dr)^2}{r^2}\right]^2\Bigg{\}} \nonumber\\
&-&\alpha \cdot  w \cdot \frac{ 4K }{r^2}\, .
\end{eqnarray}
Obviously, under the limit $D\rightarrow 4$ or $n\rightarrow 2$, the potential $V$ is regular only in the case $w=0$.   This also requires the vanished Weyl tensor of the spacetime.

However, in this case, due to the Birkhoff type theorems in four dimension, $h_{ij}$ is trivial [this tensor perturbation is based on the tensor decomposition on a two dimensional space $(N^2\, ,\gamma_{ij})$. This is very different from the gravitational perturbation based the tensor decomposition on a three dimensional space $(N^3\, ,\gamma_{ij})$, i.e.,  the case with $m=1$.].  This means the solution of the tensor perturbation equation (\ref{PE1}) is  trivial. Nevertheless, this can not hinder our
discussion on the principle symbol of  the tensor perturbation equations. Of course, to get nontrivial gravitational radiation in four dimension, one has to consider the background spacetime without the maximal symmetry of $(N^2, \gamma_{ij})$.

\section{Conclusion and discussion}
In this paper, to get a well defined principle symbol, we have shown the spacetimes in the so-called ``Einstein-Gauss-Bonnet Gravity in four dimension" have to be (locally) conformally flat. So, locally, the metric 
always has a form $g_{\mu\nu}= \Omega^2 \eta_{\mu\nu}$. Although the theory is diffeomorphism invariant when $D>4$,  the final four dimensional theory has  preferred spacetimes, and  it can not be diffeomorphism invariant. 
This point is similar to the conclusions in~\cite{Gurses:2020rxb, Gurses:2020ofy, Arrechea:2020gjw, Arrechea:2020evj}.

\section*{Acknowledgement}
This work was supported in part by the National Natural Science Foundation of China with grants
No.11622543, No.12075232, No.11947301, and No.12047502.  This work is also supported by the 
Fundamental Research Funds for the Central Universities under Grant No: WK2030000036.

\end{document}